\documentclass[reprint,aps,pra,notitlepage,superscriptaddress,nofootinbib,longbibliography,twocolumn,10pt,floatfix]{revtex4-2}
\usepackage[utf8]{inputenc}
\usepackage{amsmath,amsthm,amssymb,amsfonts}
\usepackage{mathtools}
\usepackage{graphicx}
\usepackage{makecell}
\usepackage{enumitem}
\usepackage{mathdots}
\usepackage{comment}
\usepackage[dvipsnames]{xcolor}
\usepackage{braket}
\usepackage{bbm}
\usepackage{bm}
\usepackage[title]{appendix}
\usepackage[bb=boondox]{mathalfa}
\usepackage{tikz}
\usepackage{tikz,tikz-3dplot,pgfplots}
\usetikzlibrary{arrows.meta}
\usetikzlibrary{shapes,arrows,patterns}
\usetikzlibrary{3d}
\usepackage{tikz-cd}
\newcommand{\ie}{{\it i.e.},\ }

\newcommand{\e}{\operatorname{e}}

\newcommand{\mi}{\mathrm{i}}

\pgfplotsset{compat=1.18}
\begin{document}
\title{Probing Spacetime Topology and Superposition with Accelerated Detectors}
\author{P. Poopathysankar}
\email{p.poopathysankar@gmail.com}
\affiliation{Department of Theoretical Physics, University of Madras, Chennai, India}
\author{Lucas Hackl}
\email{lucas.hackl@unimelb.edu.au}
\affiliation{School of Mathematics and Statistics, The University of Melbourne, Parkville, VIC 3010, Australia}
\affiliation{School of Physics, The University of Melbourne, Parkville, VIC 3010, Australia}
\author{Anwesha Chakraborty}
\email{anwesha.chakraborty@unimelb.edu.au}
\affiliation{School of Mathematics and Statistics, The University of Melbourne, Parkville, VIC 3010, Australia}
\begin{abstract}
We study entanglement harvested by Unruh–DeWitt detectors following Rindler trajectories in compactified and superposed Minkowski spacetime. We consider different directions of acceleration (both parallel and anti-parallel), separation between detectors and direction of spatial compactification mutually perpendicular to each other. Using the standard entanglement harvesting protocol, we analyze how these features influence the extracted correlations. When detector separation is perpendicular to the direction of acceleration, the harvested entanglement is uniformly suppressed due to increased spacelike separation. Compactification enhances field correlations leading to an increased concurrence and an extended harvesting range at higher accelerations. Additionally, we show that spacetime superposition introduces interference effects that further enlarge the entanglement harvesting region in parameter space, particularly in the high-acceleration regime. We also find that the effect of anti-parallel acceleration yielding significantly higher entanglement than parallel acceleration prevails in compactified and superposed spacetime.
\end{abstract}

\maketitle

\section{Introduction}
Relativistic quantum information theory extends the concepts and tools of conventional quantum information to regimes where relativistic effects cannot be neglected. In such settings, quantum properties such as spin, entanglement, and measurement outcomes may depend nontrivially on observer motion, causal structure, spacetime curvature, and boundary conditions. In recent years, progress in relativistic quantum information and quantum field theory in curved spacetime has clarified how central quantum notions, including entanglement, superposition, and measurement, can be consistently combined with relativistic concepts such as proper time, acceleration, horizons, and spacetime geometry~\cite{Birrell:1982ix,mann2012relativistic,Danielson:2022tdw}.

A particularly useful operational framework for exploring these questions is provided by entanglement harvesting~\cite{Mart_n_Mart_nez_2012,PhysRevD.79.044027}. It refers to the extraction of non-classical correlations from a quantum field by two initially uncorrelated localized quantum probes through their local interactions with the field. In the standard setting~\cite{PhysRevD.92.064042,PhysRevD.79.044027}, these probes are modeled as Unruh-Dewitt (UDW) detectors, namely idealized two-level quantum systems locally coupled to a scalar quantum field along prescribed worldlines. Even when the detectors remain spacelike separated, the pre-existing correlations present in the vacuum or in other field states can be transferred to the detectors during the interaction, thereby generating bipartite entanglement between them. This makes entanglement harvesting an operational probe for understanding the correlation structure of quantum fields~\cite{MartnMartnez2016,Wu2025qqu}, and it has become an important tool for studying how detector's motion, temperature, spacetime topology, and curvature affect quantum correlations.

Extensive studies have investigated entanglement harvesting in a wide range of spacetime backgrounds~\cite{chakraborty2024entanglementharvestingquantumsuperposed,e15051847,Foo2021,Liu2022,MartnMartnez2016,Liu2025ctf,PhysRevD.97.124028,Sun2021InquiringTU}, detector trajectories~\cite{PhysRevD.92.064042,Salton2015}, and switching profiles~\cite{Henderson2020}.

Among these, accelerated detector setups are of special interest because of their connection with the Unruh effect~\cite{Fuentes_Schuller_2005,Alsing_2006,Foo:2020xqn}, according to which a uniformly accelerated observer perceives the Minkowski vacuum as a thermal state~\cite{Unruh:1976db}. This observer-dependent thermal response modifies both the local noise experienced by the detectors and the nonlocal field correlations available for harvesting, thereby affecting the amount and structure of the harvested entanglement~\cite{PhysRevD.107.045010}. Early studies showed that acceleration can enhance entanglement harvesting, with uniformly accelerating detectors able to extract more entanglement than inertial ones in certain configurations, particularly when accelerating in parallel~\cite{Salton2015}. However, subsequent work demonstrated that this enhancement is not universal: acceleration can also suppress entanglement due to Unruh induced thermal noise and horizon effects, with the outcome depending sensitively on detector alignment, trajectory, and switching functions~\cite{Liu2022}. More recently, the interplay between acceleration and quantum superposition of spacetime has been explored, where the response of an accelerated detector exhibits interference between different geometric branches, indicating that acceleration interacts non-trivially with quantum spacetime structure~\cite{goel2024accelerateddetectorsuperposedspacetime}. Together, these results highlight that entanglement harvesting with accelerated detectors is governed by a subtle balance between motion-induced enhancement, thermal noise, and geometric effects.

In parallel, spacetime topology and boundary conditions are known to modify the field mode structure and, consequently, the vacuum correlations accessible to localized detectors. Compactified Minkowski spacetimes provide a simple but physically rich setting in which such topological effects can be analyzed explicitly. For inertial or static detectors, entanglement harvesting in compactified spacetimes has been studied in detail~\cite{Langlois:2005nf,MartnMartnez2016,PhysRevD.97.124028,Sun2021InquiringTU}. 

So a natural and important extension in the study would be to investigate the combined effect of acceleration and spacetime compactification on entanglement harvesting. According to~\cite{goel2024accelerateddetectorsuperposedspacetime} acceleration changes single detector's response through the Unruh effect, while compactification changes the field correlations through the altered global mode structure. So their interplay can naturally alter the harvested entanglement between two UDW detectors while they are interacting with a background quantum field, which is the primary objective in present work.
We here analyze both parallel and anti-parallel acceleration scenarios, taking the detector separation to be perpendicular to the direction of acceleration and compactification. This allows us to isolate how acceleration, relative detector motion, and spacetime topology jointly affect the  the harvested entanglement.

As a further extension, we also consider the possibility that the compactified spacetime geometry itself is placed in a quantum superposition. If spacetime is ultimately governed by quantum mechanics, then it is natural to ask whether distinct geometrical configurations may, in principle, appear in coherent superposition~\cite{PhysRevD.107.045014,Foo:2023yxj}. Recent works~\cite{Foo2021,suryaatmadja2023,chakraborty2024entanglementharvestingquantumsuperposed,Howl2022oqz,Foo_2021,Tang2025mtc,Walleghem2026rad} has indicated how quantum features of geometry can affect entanglement harvesting and quantum resources. Motivated by these developments, we investigate how the harvested entanglement is further modified when the accelerated detectors interact with a field defined on a superposition of compactified Rindler geometries.

Our analysis therefore brings together three ingredients that have so far mostly been studied separately: acceleration, spacetime compactification, and quantum superposition of geometries. While previous studies have considered accelerated entanglement harvesting~\cite{Salton2015,Liu2022}, static detectors in compactified and superposed spacetimes~\cite{MartnMartnez2016,chakraborty2024entanglementharvestingquantumsuperposed}, and the excitation probability of single accelerated detectors in superposed spacetime settings~\cite{goel2024accelerateddetectorsuperposedspacetime}, the present work investigates bipartite entanglement harvesting by accelerated detectors in compactified Rindler space and its quantum superposed extension. This provides a unified framework for examining how relativistic motion and global spacetime structure combine to shape the extraction of quantum correlations from a field.\\

The paper is organized as follows. In Sec-~\ref{sec:model}, we introduce the detector model, describe the quotient Rindler spacetime and its superposition, and derive the relevant Wightman functions and detector dynamics. In Sec-~\ref{sec:quantify}, we quantify the harvested entanglement using negativity and concurrence. In Sec-~\ref{sec:discussion}, we present and discuss our results, focusing on the roles of compactification, acceleration, and spacetime superposition. Finally, we conclude in Sec-~\ref{sec:discussion} with a summary and outlook.

\section{Model and Analysis}\label{sec:model}
In entanglement harvesting protocols, the harvested correlation is typically quantified from the reduced two detector density matrix obtained after the interaction, making Unruh-Dewitt detectors a natural and versatile framework for detecting and characterizing field mediated entanglement~\cite{PhysRevD.79.044027,Salton2015,goel2024accelerateddetectorsuperposedspacetime,chakraborty2024entanglementharvestingquantumsuperposed,MartnMartnez2016}. We use accelerated detectors to probe Rindler spacetime with non-trivial topology namely a quotient Rindler spacetime and thereafter superposition of two such Rindler spacetime. 

In this section we review the model including detector trajectories and definition of scalar field in quotient Rindler spacetime and thereafter show the analysis of joint detector state after interaction.

\subsection{Detector Trajectories and Field Interaction}
\begin{figure}[t!]
\centering
\includegraphics[width=0.8\columnwidth]{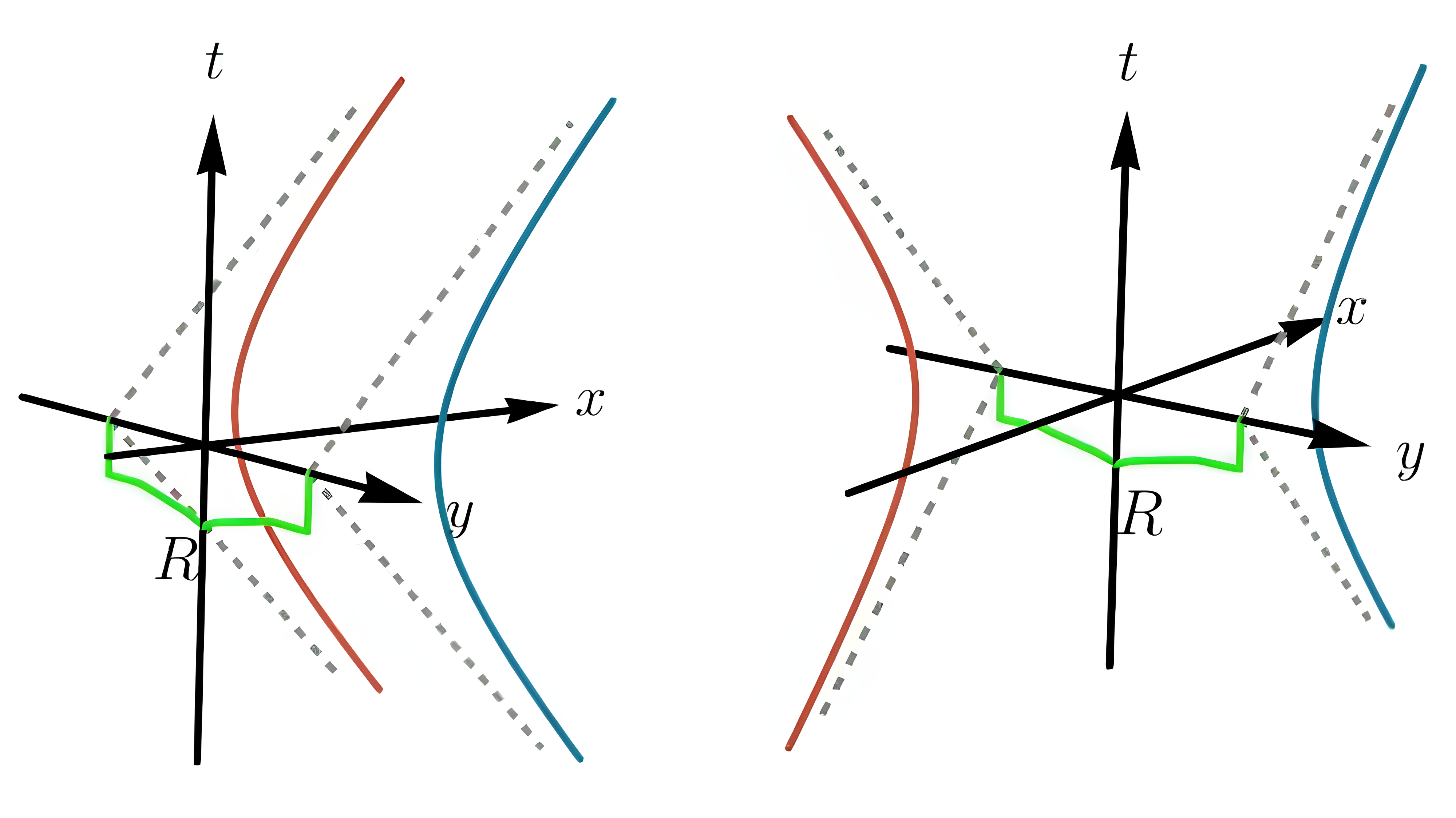}
\caption{\textit{Accelerated detector trajectories.} Here we represent three out of the four dimensions of Minkowski space
$(x,y,t)$ and the detector separation in the $y$-direction.
The blue curves represent the trajectories of detector A while the red curve
represents the trajectory of detector B.
The left panel represents the parallel acceleration case while the right panel
represents the anti-parallel acceleration case.}
\label{fig:detector}
\end{figure}

We model the detectors $D$ as point-like two level quantum systems, each characterized by an energy gap $\Omega_D$, and assume that they interact with the field through an UDW coupling. The detectors are separated along the $y$-axis and uniformly move with either parallel or anti-parallel acceleration along the $x$-axis as shown in Fig~\ref{fig:detector}.  
%we can write the transformation between Minkowski $(t,x,y,z)$ and Rindler coordinate $(\tau,\frac{1}{a},y,z)$ as \begin{equation}t = \frac{1}{a}\sinh{(a\tau)},\,\,x = \frac{1}{a} \cosh{(a\tau)},\,\,y=y,\,\,z=z\end{equation}
The Rindler coordinates of the detectors $A$ and $B$ in (anti-)parallel acceleration are respectively given by
\begin{align}
    t_A&=\frac{1}{a}\sinh(a\tau), &x_A&=\frac{1}{a}\cosh(a\tau), &y_A&=-\frac{R}{2}\nonumber,\\
    t_B&=\frac{1}{a}\sinh(a\tau'), &x_B&=\pm\frac{1}{a}\cosh(a\tau'), &y_B&=\frac{R}{2},\label{trajectory}
\end{align}
where $a$ is the proper acceleration and $\tau$ is the proper time of the accelerated detectors. Both of them are at the origin of the $z$-axis and the sign ($\pm$) in $x_B$ denotes parallel and anti-parallel acceleration respectively.

The Hamiltonian of the detectors is given by
\begin{equation}
   H_{0}^D = \frac{\Omega_D}{2} (|e\rangle\langle e| - |g\rangle\langle g|) = \frac{\Omega_D}{2} \sigma_z,\quad D=A,B
\end{equation}
where $\sigma_z$ is respective the Pauli matrix and $\Omega_D$ is the energy gap between the excited state $|e\rangle$ and the ground state $|g\rangle$ of the respective detectors. We also consider a scalar field $\hat{\phi}$ which linearly interacts with the qubits and the interaction Hamiltonian of the detector-field system is of the form
\begin{equation}
    V_I^D(\tau)= \eta_D(\tau)\Big(\ket{e}\bra{g}+\ket{g}\bra{e}\Big)\hat{\phi}(x_D(\tau))
\end{equation}
where $\tau$ is the proper time. The function $\eta_D(\tau)$ is the switching function for the detectors specifying the interaction window between detector and field. We choose the switching to remain very small at all times, $0<\eta(\tau)\ll 1$, and to have appreciable support only over a narrow range of proper time. This corresponds to a short and weak interaction with the field. The short duration ensures that the detection events are approximately spacelike separated, while the weak coupling justifies a perturbative treatment of the interaction. We take the function to be a Gaussian with standard deviation $\sigma_D$, such that
\begin{equation}
    \eta_D(\tau) = \eta_0 e^{-\tau^2 / 2\sigma_D^2}\,,
\label{eq:switchingfunction}
\end{equation}
where $\eta_0$ is the coupling constant. Since the switching function is analytic, it cannot have strictly compact support in time. Therefore, although the detectors may be arranged so that their central interaction regions are spacelike separated for the chosen values of $\tau$, the exponentially small tails of the switching functions imply that there can still be a small temporal overlap between the detector–field couplings. In this sense, the detectors are only approximately spacelike separated. Importantly, however, the detectors do not interact directly with one another; any correlation between them is mediated by the quantum field through the interaction Hamiltonian. Consequently, the extracted correlations may contain not only contributions from pre-existing vacuum entanglement of the field, but also, in principle, small causal contributions arising from the non-compact tails of the switching functions. In the interaction picture the interaction Hamiltonian is
\begin{align}
    H_I^D&=\e^{\mi H_0^D \tau}V_{I}^D\e^{-\mi H_0^D \tau}\nonumber\\
    &= \eta_D(\tau) \big(e^{\mi\Omega_k\tau} \sigma^+ + e^{-\mi\Omega_k\tau} \sigma^-\big)\hat{\phi}(x_D(\tau)),
    \label{intera}
\end{align}
where $\sigma^+ = |e\rangle\langle g|$ and $\sigma^- = |g\rangle\langle e|$ are raising and lowering operator respectively.

\subsection{Quotient Rindler Space, its superposition  and Wightman's function}
Since we want to study the entanglement harvesting protocol in a superposed quotient space we here review the construction of such space and definition of corresponding scalar field on it. A nontrivial topological modification of Minkowski spacetime described in Rindler coordinates is given by a cylindrical spacetime $R_{0}=\mathcal{R}/J_{0}$ by quotienting the Rindler wedge $\mathcal{R}$ under the discrete isometry group $J_{0_L} : (t, x, y, z) \mapsto (t, x, y, z + L)$,
with $L$ denoting the compactification scale. Since this action is free, the resulting spacetime $\mathcal{R}_{0}$ remains a well defined Lorentzian manifold that is both space and time orientable. To formulate quantum field theory on this geometry, we introduce an automorphic field $\hat{\Phi}^L([x])$ obtained from the massless scalar field $\hat{\phi}(x)$ through an image sum over the quotient images as follows~\cite{Banach:1979iy,Banach:1979wz}:
\begin{equation}
\hat{\Phi}^L([x])=\frac{1}{\sqrt{\mathcal{N}}}\sum_{n=-\infty}^{\infty} \gamma^n \hat{\phi}(J_{0_L}^n x),\label{automorphic field}
\end{equation}
where $[x] \in \mathcal{R}_0$ corresponds to the equivalence class of $x \in \mathcal{R}$ and $\gamma$ can take values $-1$ and $1$ representing twisted and untwisted fields respectively. The normalization factor is given by $\mathcal{N} = \sum_{n=-\infty}^{\infty} \gamma^{2n}$. \footnote{We shall only study the effects with untwisted fields and use $\gamma=1$ in our calculation later on.}

The two-point correlation function, also known the Wightman function in terms of the geodesic distance in Rindler coordinatization can be written as 
\begin{align}
W_{\mathcal{R}_0}&=\bra{0}\hat{\Phi}^L([x])\hat{\Phi}^L([x'])\ket{0}\nonumber\\
    &= \frac{1}{\mathcal{N}}\sum_{n,m} \gamma^{n+m}\bra{0}\,\hat{\phi}(J^n_{0_{L}}x)\,\hat{\phi}(J^m_{0_{L}}x')\,\ket{0}\nonumber\\
    &=\frac{1}{\mathcal{N}}\sum_{n,m} \gamma^{n+m}~ W_{\mathcal{R}}(J^n_{0_{L}}x,J^m_{0_{L}}x')
\end{align}
where the usual Wightman function $W_{\mathcal{R}}(x,x')$ is
\begin{align} 
        &W_{\mathcal{R}}(x,x') =\bra{0}\hat{\phi}(x)\hat{\phi}(x')\ket{0}\nonumber\\
        &=\frac{1}{4\pi i}\text{sgn}(\tau-\tau')\delta(\sigma_{\mathcal{R}}(x,x'))-\frac{1}{4\pi^2 \sigma_{\mathcal{R}}(x,x')}.
    \label{wigh}
\end{align}
Here, the geodesic distance $\sigma_{\mathcal{R}}$ in Rindler coordinatization for parallel trajectory  is  given by
\begin{equation}
\sigma_{((}(x,x') = \frac{4}{a^2}\sinh^2\left(\frac{a}{2}(\tau-\tau') \right) -(y-y')^2 -(z-z')^2,
\label{GEO}
\end{equation}
and for equal anti-parallel acceleration 
\begin{equation}
\sigma_{)(}\, (x , x') = \frac{-4}{a^2}\cosh^2\left(\frac{a}{2}(\tau+\tau') \right) -(y-y')^2 -(z-z')^2.
\label{GEO2}
\end{equation}
The geodesic interval in~\eqref{GEO2} is relevant only for correlation functions evaluated between the two detectors when they follow anti-parallel trajectories. For quantities associated with the motion of a single detector alone \textit{i.e.} transition probability, the relevant geodesic interval remains the same as in the parallel acceleration case, since it depends only on the individual uniformly accelerated worldline.

Having reviewed the quotient Rindler spacetime corresponding to a fixed
compactification scale $L$, we now describe how this construction is extended
when the compactification scale is treated as a branch label in a quantum
superposition of geometries. Following the framework of~\cite{FoosuperMinkowski}, we associate two compactification scales $L_1$ and $L_2$
with orthogonal control states $\ket{L_1}$ and $\ket{L_2}$, which constitutes a two dimensional  spacetime Hilbert space $\mathcal{H}_S=\text{span}\{\ket{L_1},\ket{L_2}\}$. The control degree of freedom therefore labels the spacetime branch on which the field is defined, where each branch $\ket{L_i}$ corresponds to the quotient spacetime
$\mathcal{R}_{0}^{(L_i)}=\mathcal{R}/J_{0_{L_i}}$. A
superposed quotient geometry is then described at the level of the control
system by states of the form 
$\ket{s}=\sum_i c_i \ket{L_i}$.

In this setting, one must also consider the branch-dependent two-point function between the
automorphic fields defined with compactification scales $L_i$.
This motivates the definition of the Wightman function $W^{L_1L_2}(x,x')$ between any two fields, $\hat{\Phi}^{L_1}(x)$ and $\hat{\Phi}^{L_2} (x')$ is given by~\cite{FoosuperMinkowski}
\begin{align}
&W_{L_1L_2}(x,x')=\bra{0}\hat{\Phi}^{L_1}(x)\hat{\Phi}^{L_2}(x')\ket{0}\nonumber\\
&\hspace{0.6cm}=\frac{1}{\mathcal{N}}\sum_{n,m}\gamma^{n+m} \bra{0}\,\hat{\phi}(J^n_{0_{L_1}}x)\,\hat{\phi}(J^m_{0_{L_2}}x')\ket{0}\nonumber\\
&\hspace{0.6cm}=\frac{1}{\mathcal{N}}\sum_{n,m}\gamma^{n+m} W_{\mathcal{R}}(J_{0_{L_1}}x, J_{0_{L_2}}x')
\end{align}
The Wightman function above will be calculated using~\eqref{wigh} where the geodesic distance for example will be modified as
\begin{align}
    \sigma_{((}(J_{0_{L_1}}x, J_{0_{L_2}}x')&= \frac{4}{a^2}\sinh^2\left(\frac{a}{2}(\tau-\tau') \right) -(y-y')^2 \nonumber\\
    &-(z-z'+nL_1-mL_2)^2
\end{align}
in case of parallel acceleration. The Wightman's functions will be important in the calculation of the component of the joint detector's density matrix as we will see in the ensuing subsection.

\subsection{Initial State and Evolution}

The total system consists of two UDW detectors, quantum field background and the spacetime control qubit. We prepare the initial state of the detectors and field in their respective ground states $\ket{g}$ and $\ket{0}_F$ and the spacetime in an arbitrary superposition of the two spacetime branches~\cite{chakraborty2024entanglementharvestingquantumsuperposed,FoosuperMinkowski,PhysRevResearch.3.043056,goel2024accelerateddetectorsuperposedspacetime} given by
\begin{equation}
    \left|s_i\right\rangle = \cos\theta\, \left|L_1\right\rangle + \sin\theta\, \left|L_2\right\rangle. \label{s_i}
\end{equation}
So the initial state of the full system is given by
\begin{align}
    \left|\psi_i\right\rangle = \left|g,g\right\rangle \otimes \left|0\right\rangle_F \otimes \left|s_i\right\rangle, \label{initial state}
\end{align}
where $\left|g,g\right\rangle \equiv \left|g\right\rangle_A \otimes \left|g\right\rangle_B$, is the joint ground state of the detectors and $\left|0\right\rangle_F$ is the Minkowski vacuum. The unitary evolution operator in the interaction picture is given by
\begin{align}
  U&\!=e^{-\mathrm{i}\hat{\mathcal{T}}\int \, dt\, \left( \frac{d\tau_A}{dt} \hat{V}_I(\tau_A(t)) \otimes \mathbb{1}  + \mathbb{1} \otimes \frac{d\tau_B}{dt} \hat{V}_I(\tau_B(t)) \right)}\,. \label{unitary}
\end{align}

\normalsize
After time evolution the final state is
\begin{equation}
    \left|\psi_f\right\rangle = \sum_n \lambda^n \left|\psi_f^{(n)}\right\rangle = \sum_n U_n \left|\psi_i\right\rangle, \label{finalstate}
\end{equation}
where $U_n$ is the term in $U$ of order $\lambda^n$. We then condition on the final spacetime state being in the state
\begin{equation}
    \left|s_f\right\rangle = \cos\phi\, \left|L_1\right\rangle + \sin\phi\, \left|L_2\right\rangle. \label{eq:s_f}
\end{equation}
Conditioning on a final spacetime state is essential if one wants to retain operational information about the coherence between different spacetime branches. If the spacetime control degree of freedom is simply traced out, the resulting detector state becomes an incoherent mixture of the contributions from the individual geometries, and the interference between distinct compactification scales are lost. By contrast, post-selecting or conditioning on a final spacetime state projects the joint detector-spacetime state onto a chosen coherent branch superposition, thereby allowing the reduced detector density matrix to contain cross terms between different geometries. This procedure is therefore crucial for probing genuinely quantum features of the spacetime superposition through detector observables, since it makes the harvested correlations sensitive not only to each classical geometry separately, but also to the relative phases between the spacetime branches.

The reduced state of the detectors is then
\begin{equation}
    \rho_{AB} = \mathrm{Tr}_{\Phi} \left[ \left\langle s_f \middle| U \middle| \psi_i \right\rangle \left\langle \psi_i \middle| U^\dagger \middle| s_f \right\rangle \right],
\end{equation}
which yields a density operator of $X$ form
\begin{equation}
    \rho_{AB} = \begin{pmatrix}
        P^G & 0 & 0 & X \\
        0 & P^{E}_B & C & 0 \\
        0 & C^* & P^{E}_A & 0 \\
        X^* & 0 & 0 & 0
    \end{pmatrix}. \label{density matrix}
\end{equation}
in the basis $\left\{ \left|0,0\right\rangle, \left|0,1\right\rangle, \left|1,0\right\rangle, \left|1,1\right\rangle \right\}$. Here, we have used perturbative analysis and constructed the density matrix only up to order of $\lambda^2$~\cite{PhysRevD.79.044027,Salton2015,MartnMartnez2016}. The explicit matrix elements are given by
\small
\begin{align}
    &P^G=(a+b)^2-\lambda^2\Big[(a^2+ab)\sum_{D}P_D^{L_1}+(b^2+ab)\sum_{D}P_D^{L_2}\Big]\label{pground}\\
    &P^{E}_{D}=\lambda^2(a^2~P_D^{L_1}+b^2~P_D^{L_2}+2a b~P_D^{L_1L_2}),\label{PDE}\\ 
   &C= \lambda^2\int dt\,dt' \nu_A(t)\bar{\nu}_B(t')\Big[a^2~W^{L_1}\Big(x_A(t),x_B(t')\Big)\nonumber\\
   &+b^2 ~W^{L_2}\Big(x_A(t),x_B(t')\Big)+2ab  ~W^{L_1L_2}\Big(x_A(t),x_B(t')\Big)\Big],\label{c term}\\
   &X=-\lambda^2\int dt\,dt'\nu_A(t)\nu_B(t')\Big[(a^2+ab)W^{L_1}(x_B(t'),x_A(t))\nonumber\\
   &\hspace{80pt}+(b^2+ab)~W^{L_2}(x_B(t'),x_A(t))\Big],\label{X}
\end{align}
\normalsize
with $\nu_D(t)=\chi_D(t)\e^{-\mi \Omega_D\tau_D(t)}$ and $a 
= \cos\theta \cos\phi$ and $b = \sin\theta \sin\phi$. 
Additionally, $P_D^{L_i}$ is individual detector's transition probability in single compactified spacetime with length $L_i$ and given by 
\begin{align}
   &P_D^{L_i}= \int dt\,dt'\nu_D(t)\bar{\nu}_D(t')\,W^{L_i}\Big(x_D(t),x_D(t')\Big),\label{PDLi}
\end{align}
On the other hand $P_D^{L_1L_2}$ is an interference term contributing to transition probability arising as an effect of spacetime superposition and is explicitly given by
\begin{align}
  & P^{L_1L_2}_D=\int dt\,dt' \nu_D(t)\bar{\nu}_D(t')\,W^{L_1L_2}\Big(x_D(t'),x_D(t)\Big).\label{PDL1L2}
  \end{align}
In ensuing section we discuss the conditions for entanglement generation and quantify harvested entanglement using these matrix components and further discuss our findings.

\section{Quantifying Entanglement Harvesting}\label{sec:quantify}

To give a qualitative discussion of the effect of acceleration, spatial compactification and geometry superposition on the harvested entanglement, we quantify the entanglement using standard entanglement measures for two-qubit systems --- negativity and concurrence. 

Negativity is based on the Peres--Horodecki (PPT) criterion~\cite{PhysRevLett.77.1413,liu2026trianglecriterionmixedstatemagic,Vidal_2002}. For a density matrix $\rho_{AB}$, the partial transpose 
with respect to subsystem $A$ is denoted by $\rho^{\Gamma_A}$. For two-qubit systems, a state is entangled if and only if 
$\rho^{\Gamma_A}$ possesses at least one negative eigenvalue and eventually produce positive value of the following quantity
\begin{equation}
\mathcal{N}=\frac{\|\rho^{\Gamma_A}\|_1 - 1}{2},
\end{equation} 
known as \textit{entanglement negativity.} For our reduced density matrix, the PPT criterion leads to two possible conditions so that $\mathcal{N}$ has nonzero positive value. However, within the perturbative regime considered here, only one of them can be satisfied, yielding the entanglement condition
\begin{equation}
|X| > P_D^E,
\label{condition}
\end{equation}
while the other condition will always analytically fail.
So the negativity simplifies to
\begin{equation}
\mathcal{N}=\max\!\left(0,\, |X| - P_D^E \right).
\end{equation}
Concurrence was introduced by Wootters (1998)~\cite{Wootters:1997id,Rungta_2001} as a closed-form entanglement measure for mixed two-qubit states and is directly related to the entanglement of formation. For two identical detectors ($\Omega_A=\Omega_B,\,\eta_A=\eta_B$), concurrence is directly proportional to the negativity,
\begin{equation}
\mathcal{C}=2\mathcal{N}=\max\!\Big(0,\,2(|X|-P^E)\Big).
\end{equation}
Entanglement can thus be understood as arising from a balance between the local excitation probability of each detector $P^E$ and the off diagonal coherence term $X$ in the reduced density matrix, which encodes the amplitude for field mediated exchange process. When the interaction regions of the detectors are strictly spacelike separated, this exchange contribution cannot be attributed to the causal transmission of real excitations through the field. Instead, it is more appropriately interpreted as the extraction of pre-existing entanglement from the field and its transfer to the detectors.

\subsection{Local detector excitation probabilities}

The local excitation probabilities describe the noise experienced by each detector through its interaction with the field. Since this is a local quantity, depending only on the pullback $W(x_D(\tau),x_D(\tau'))$ along a single detector worldline, it is insensitive to whether the other detector moves parallel or anti-parallel. Therefore, it can be computed using the same local geodesic interval, namely~\eqref{GEO}. 

The excitation probability for detector $D\in\{A,B\}$ takes the form
\begin{equation}
P^{E}_{D}=\lambda^2\left(a^2 P_D^{L_1}+b^2 P_D^{L_2}+2ab P_D^{L_1L_2}\right),
\end{equation}
where $P_D^{L_i}$ represents the contribution from a single compactified branch, while $P_D^{L_1L_2}$ represents interference between the two branches arising from the coherent superposition of geometries. The explicit expressions for $P_D^{L_i}$ and $P_D^{L_1L_2}$ are  given by
\begin{widetext}
    \begin{align}
        P_D^{L_i} &= \sum_{m=n}\frac{\sqrt{\pi}\sigma}{\mathcal{N}}\frac{\eta_0^2}{4\pi} \left[ -\Omega -\frac{a^2}{2\pi} \int_{0}^{\infty} dv\,\left(  \frac{e^{-\frac{v^2}{4\sigma^2}} \cos(\Omega v)}{\sinh^2(\frac{a}{2}v)}  - \frac{4}{a^2 v^2}\right) \right]-\sum_{m\ne n}\frac{\eta_0^2 \sigma}{2\mathcal{N} \sqrt{\pi} }\Bigg[\frac{\e^{\frac{-d_i^2}{ 4\sigma^2}}  \sin{(\Omega d_i)}}{(m-n)L_i\sqrt{1+\frac{a^2((m-n)L_i)^2}{4}}}\nonumber\\
        &+ \frac{1}{2\pi}   \int_{-\infty}^{\infty} dv \, \frac{e^{\frac{-v^2}{ 4\sigma^2}}  e^{-i\Omega v} }{\left(\frac{4}{a^2}\sinh^2\left(\frac{a}{2}v \right)-((m-n)L_i)^2\right)}\Bigg]\label{PDLifinal}
    \end{align}
    \begin{align}
        P_D^{L_1L_2} &= \sum_{m,n=0}\frac{\sqrt{\pi}\sigma}{\mathcal{N}}\frac{\eta_0^2}{4\pi} \left[ -\Omega -\frac{a^2}{2\pi} \int_{0}^{\infty} dv\,\left(\frac{e^{-\frac{v^2}{4\sigma^2}} \cos(\Omega v)}{\sinh^2(\frac{a}{2}v)}  - \frac{4}{a^2 v^2}\right) \right] -\sum_{m,n\ne0}\frac{\eta_0^2 \sigma}{2\mathcal{N} \sqrt{\pi}}\Bigg[\frac{e^{\frac{-d_K^2}{ 4\sigma^2}}  \sin{(\Omega d_K)}}{K_{L_1L_2}\sqrt{1+\frac{a^2(K_{L_1L_2})^2}{4}}}\nonumber \\ 
        & + \frac{1}{2 \pi}  \int_{-\infty}^{\infty} dv \, \frac{e^{\frac{-v^2}{ 4\sigma^2}}  e^{-i\Omega v}}{\left(\frac{4}{a^2}\sinh^2\left(\frac{a}{2}v \right)-K_{L_1L_2}^2\right)}\Bigg]\label{PDL1L2final}
    \end{align}
where $d_i=\frac{2}{a}\sinh^{-1}\left(\frac{a}{2}(m-n)L_i\right)$, $d_{K}=\frac{2}{a}\sinh^{-1}\left(\frac{aK_{L_1L_2}}{2}\right)$ and $K_{L_1L_2}^2=(nL_1-mL_2)^2$.
\end{widetext}

\subsection{Nonlocal correlation terms}
In contrast to the transition probability, the $X$ term encodes nonlocal correlations between the two detectors and depends on the mixed pullback $W(x_A(\tau),x_B(\tau'))$, which explicitly involves both trajectories simultaneously. As a result, the spacetime interval entering $X$ changes between the parallel~\eqref{GEO} and anti-parallel~\eqref{GEO2} configurations, leading to different Wightman functions and thus different forms $X_{((}$ and $X_{)(}$ respectively as shown below.
\begin{widetext}
\begin{align}
    &X_{((} = -2 \lambda^2 \eta_0^2 \sqrt{\pi}\,\sigma \, e^{-\sigma^{2}\Omega^{2}} \sum_{m,n=-\infty}^{\infty}\Bigg[ (a^2 + ab)  \Bigg(- \frac{\exp{\Big[\frac{-\left(\sinh^{-1}\left(\frac{D_1a}{2}\right)\right)^2}{a^2\sigma^2}}\Big]}{8\pi iD_1 \,\, \sqrt{1+\left(\frac{D_1a}{2}\right)^2}}  - \frac{1}{{4\pi^2}}\int_{-\infty}^{0} dv \,   \frac{e^{-v^2 / 4\sigma^2}}{\left(\frac{4}{a^2}\sinh^2\left(\frac{av}{2} \right)-D_1^2\right)}\Bigg) \nonumber \\
    &+ (b^2 + ab) \Bigg(- \frac{\exp{\Big[\frac{-\left(\sinh^{-1}\left(\frac{D_2a}{2}\right)\right)^2}{a^2\sigma^2}}\Big]}{8\pi iD_2 \,\, \sqrt{1+\left(\frac{D_2a}{2}\right)^2}}  - \frac{1}{{4\pi^2}}\int_{-\infty}^{0} dv \,   \frac{e^{-v^2 / 4\sigma^2}}{\left(\frac{4}{a^2}\sinh^2\left(\frac{av}{2} \right)-D_2^2\right)}\Bigg) \Bigg],\label{Xpara}\\
    &X_{)(} = -\frac{\lambda^2\eta_0^2\sigma\sqrt{\pi}}{{8\pi^2}\mathcal{N}}   \sum_{m,n=-\infty}^{\infty}\Bigg[ (a^2 + ab) \int_{-\infty}^{\infty} du  \, \frac{e^{-u^2 / 4\sigma^2}  e^{-i\Omega u}}{{\frac{4}{a^2}\cosh^2\left(\frac{a}{2}(u) \right)+D_1^2}} + (b^2 + ab)\int_{-\infty}^{\infty} du  \, \frac{e^{-u^2 / 4\sigma^2}  e^{-i\Omega u}}{{\frac{4}{a^2}\cosh^2\left(\frac{a}{2}(u) \right)+D_2^2}} \Bigg],\label{xantipara}\\
    &\text{where $D_i=\sqrt{R^2+(m-n)^2L_i^2}$ \quad for $i=1,2$ .}\nonumber
\end{align}
The full derivation, including intermediate steps and explicit evaluation of the Wightman integrals, is provided in the Appendix-\ref{App:A} and~\ref{App:B}.
\end{widetext}

\begin{figure*}[t]
\centering
\begin{tikzpicture}
  \node[anchor=south west, inner sep=0] (img)
    at (0,0) {\includegraphics[
      width=0.82\textwidth,
      height=0.6\textheight,
      keepaspectratio
    ]{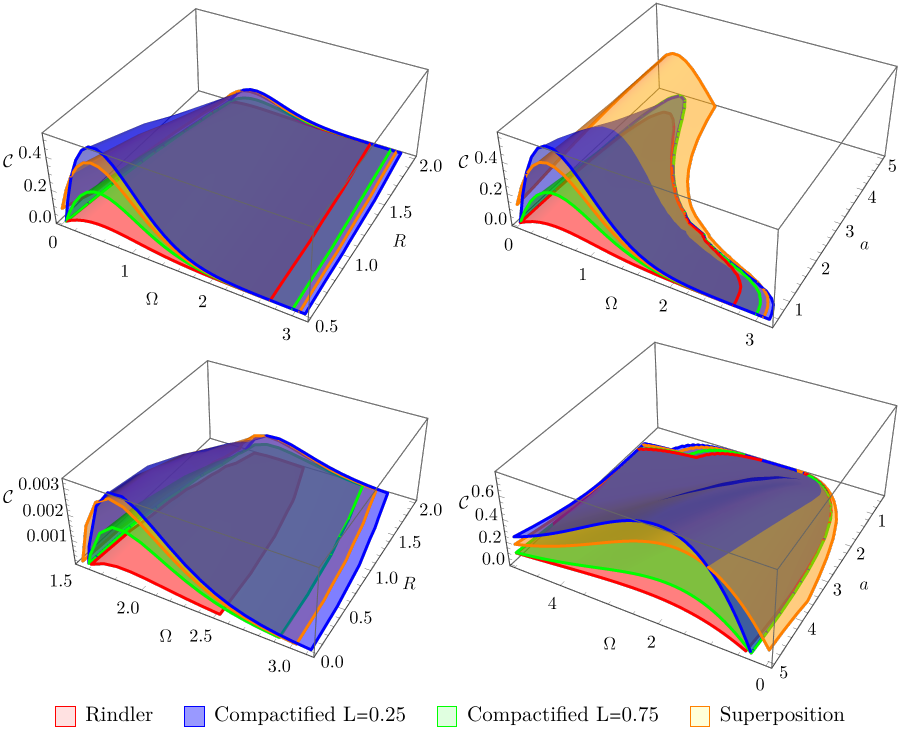}};

  \begin{scope}[x={(img.south east)}, y={(img.north west)}]
    \node at (0.2,0.55) {(a)};
    \node at (0.7,0.55) {(b)};
    \node at (0.2,0.08) {(c)};
    \node at (0.7,0.08) {(d)};
  \end{scope}
\end{tikzpicture}
\caption{\textit{Concurrence $\mathcal{C}$ for accelerated detectors in various spacetime geometry}. Panels (a) and (b) show the parallel acceleration configuration, while panels (c) and (d) show the anti-parallel configuration. The left column presents $\mathcal{C}$ as a function of the energy gap $\Omega$ and detector separation $R$ for constant acceleration $a=0.5$, whereas the right column presents $\mathcal{C}$ as a function of $\Omega$ and acceleration $a$ for constant detector separation $R=0.2$. All figures are shown for four spacetime configurations \ie Rindler spacetime, compactified Rindler spacetime with $L=0.25$ and $L=0.75$, and the symmetric superposition of the two compactified branches. The plots show that both range and region of entanglement is substantially increased for compactified space in compared to usual Rindler spacetime where as with symmetric superposed compactified spacetime the region of entanglement is enlarged particularly along the direction of acceleration. Parallel and anti-parallel acceleration show contrasting behavior as discussed in subsection-\ref{sec:acceleration}.}
\label{fig:concurrence}
\end{figure*}
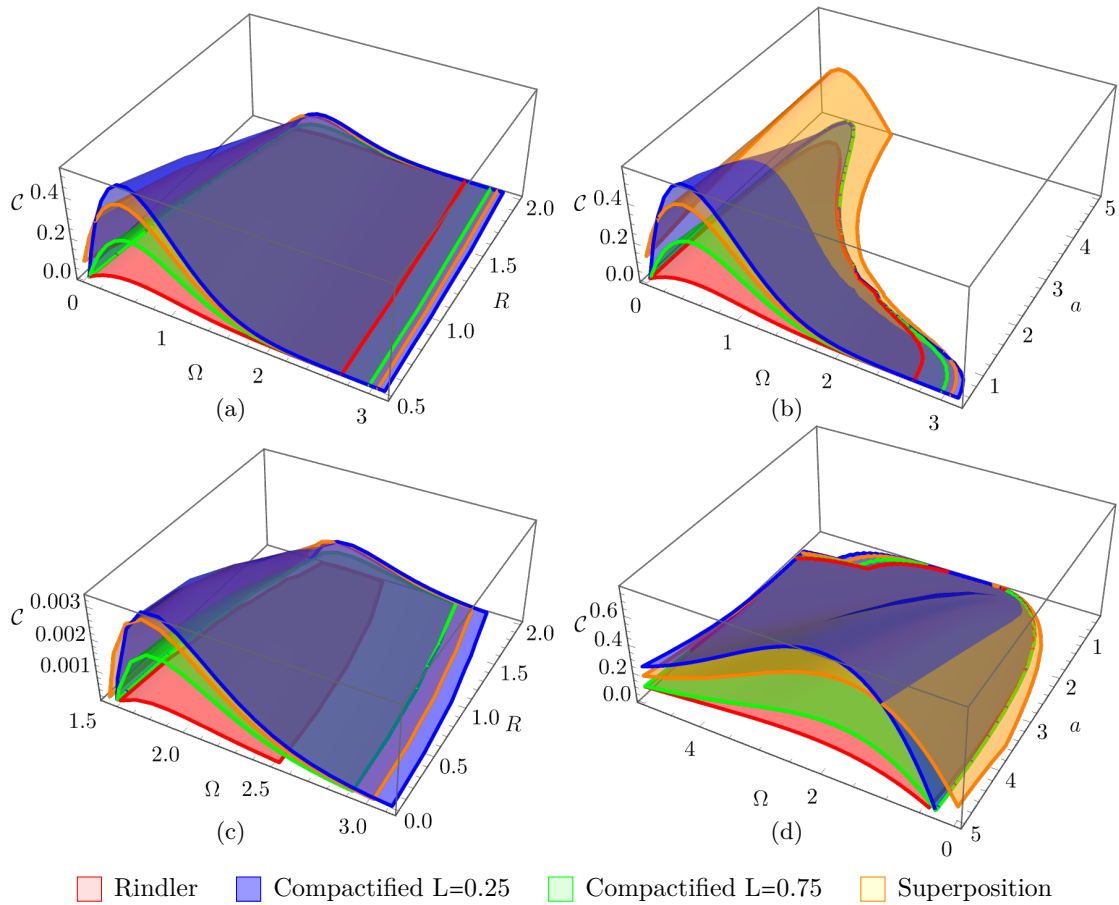

\section{Discussion: Range and Robustness of Entanglement Harvesting}\label{sec:discussion}

This work addresses how acceleration, compactification, and coherent superposition of geometries jointly affect entanglement harvesting between two UDW detectors. We plotted the concurrence over two parameter sets, $\mathcal{C}(R,\Omega)$ and $\mathcal{C}(a,\Omega)$, with $R,\Omega,a$ being the separation between detectors along $y$ axis, energy gap and acceleration of the detectors respectively. For both parallel and anti-parallel acceleration, we have plotted the entanglement properties in fig:~\ref{fig:concurrence} for four spacetime settings: a Rindler space,  compactified Rindler space with two separate compactification lengths $L=0.25$ and $L=0.75$, and a symmetric superposition of the two compactified spacetime sectors\footnote{In a previous study~\cite{chakraborty2024entanglementharvestingquantumsuperposed} we already established that the effect of superposed spacetime is most pronounced when $\theta=\phi$. We have chosen $\theta=\phi=\frac{\pi}{4}$ for our analysis.} . Across these studies we find three robust trends:
\begin{enumerate}
\item an universal suppression in harvested entanglement due to the chosen initial separation between two detectors which is \emph{perpendicular} to the direction of acceleration.
\item a systematic enhancement and expanded harvesting region due to compactification of spacetime.
\item superposed geometry enlarges the entanglement harvesting region, particularly in the high acceleration regime.
\end{enumerate}

Apart from this we see a general trend of enhance concurrence value and entanglement region for anti-parallel acceleration in contrast to parallel trajectory specifically for higher acceleration region. This feature has already been discussed in previous literature~\cite{Salton2015,Liu2022,PhysRevD.107.045010}. We discover the same feature persisting for compactified and superposed geometry. In ensuing subsections we do quantitative and qualitative discussions of these individual features.

\subsection{Perpendicular separation: entanglement suppression due to geometry}

As shown in figure~\ref{fig:concurrence}(a) and (c) concurrence peaks at a specific positive $\Omega$ values of detector's energy gap. For short interaction time, the Gaussian switching has a broad Fourier profile, so the detector samples a wider range of field frequencies rather than sharply selecting a single resonant mode. The concurrence peak therefore appears as a broad Gaussian-like feature. Additionally, the peak decreases when detector's initial separation $R$ is increased. This is quite intuitive and has been studied for various geometry and trajectories in literature. 
However, interestingly, in our case, the detectors accelerate along the $x$-direction while their mutual separation is fixed along the transverse $y$-direction. This geometry produces an additional, unavoidable contribution to the spatial interval entering the Wightman function. The relevant geodesic separation in \eqref{GEO} and~\eqref{GEO2} contains a nonzero $(\Delta y)^2$ term on top of the acceleration-induced growth in the $x$-separation. As a consequence, the detectors become effectively more spacelike separated throughout the switching window, which suppresses the Wightman kernel and reduces the nonlocal correlation term $X$.

\begin{figure}[!t]
    \centering
    \includegraphics[
      width=0.75\columnwidth,
      height=0.32\textheight,
      keepaspectratio
    ]{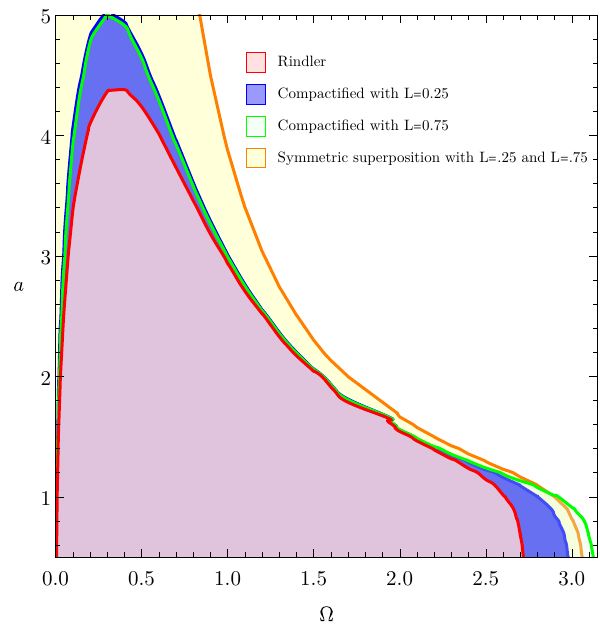}
    \caption{\textit{Region of Concurrence in the $(\Omega,a)$ parameter space.} The shaded regions indicate the parameter domains in which the concurrence $\mathcal{C}$ is nonzero for different spacetime configurations for the case of parallel acceleration corresponding to figure\ref{fig:concurrence}b. The entanglement region is larger in compactified Rindler spacetime (blue and green) than in ordinary Rindler spacetime (red), and is further enlarged in the symmetric superposition of the two compactified Rindler branches (yellow).}
    \label{fig:area}
\end{figure}

\begin{quote}
This transverse geometry acts as a \emph{geometric penalty} which lowers the baseline harvesting level against which the effects of compactification and superposition are assessed afterwards.
\end{quote}

Numerically, this manifests as a \emph{uniform decrease} of concurrence as we increase $R$, relative to otherwise comparable harvesting scenarios studied in the literature~\cite{Liu2022,Salton2015,PhysRevD.107.045010} where the separation is along the \textit{same direction} of acceleration. Importantly, this effect is geometry independent and shows up in all four cases.  

\begin{figure*}[tbp]
\centering
\resizebox{.8\textwidth}{!}{
\begin{tikzpicture}
  \node[anchor=south west, inner sep=0] (img) 
    at (0,0) {\includegraphics[width=\linewidth]{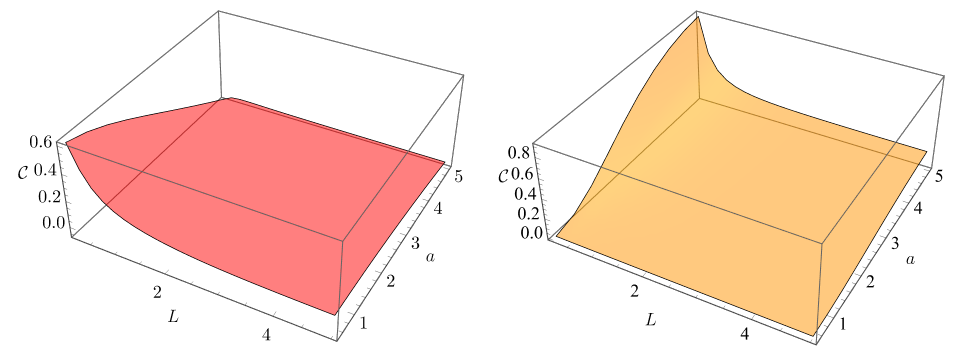}};

  \begin{scope}[x={(img.south east)}, y={(img.north west)}]
    \node at (0.2,0.0) {(a)};
    \node at (0.7,0.0) {(b)};
  \end{scope}
\end{tikzpicture}}
\caption{\textit{Concurrence $\mathcal{C}$ as a function of compactification length $L$  and acceleration $a$.} The plots illustrate the combined effect of acceleration and compactification. (a) corresponds to parallel acceleration showing the higher concurrence value for small compactification length and low acceleration which becomes progressively weaker at higher compactification length and acceleration. (b) corresponds to anti-parallel acceleration where concurrence is larger for small $L$ but high value of acceleration in comparison to parallel acceleration scenario.}
\label{fig:con_acc_L}
\end{figure*}

\subsection{Compactification enhanced entanglement and its interplay with acceleration}

Compactifying one spatial direction modifies the Wightman function through an image sum in the field, introducing additional correlation compared to uncompactified (Rindler) case. In the $C(a,\Omega)$ plot in figure~\ref{fig:concurrence}(b) and (d), this produces two clear effects relative to the Rindler case: (i) value of concurrence increases in sizable regions of parameter space, as seen in figure~\ref{fig:concurrence}(b)and (d); and (ii) harvesting persists for larger acceleration range giving rise to larger region of entanglement as seen in figure~\ref{fig:area}.

In our numerical range, the uncompactified Rindler configuration ceases to harvest beyond approximately $a\simeq 4.4$, whereas both compactified branches retain nonzero concurrence up to approximately $a\simeq 5$ for $0<\Omega< 2$. Also, the peaks for compactified geometry (green and blue) is noticeably higher than that of Rindler geometry (red) for both parallel and anti-parallel trajectories.
\begin{quote}
	This indicates that compactification does not merely enhances the harvested entanglement, it also extends the acceleration range over which entanglement can be extracted. 
\end{quote}

We also notice the concurrence peak shifting along the $\Omega$-axis can be understood
from the acceleration dependence of the field correlations sampled by the
detectors. In the detector response integrals, the phase factor
$e^{i\Omega \tau}$ is weighted by the Wightman function evaluated along the
accelerated trajectories. As the acceleration $a$ changes, the geodesic
interval entering the Wightman function changes through factors such as
\[
\sinh^2\!\left[\frac{a}{2}(\tau-\tau')\right]
\quad \text{or} \quad
\cosh^2\!\left[\frac{a}{2}(\tau+\tau')\right],
\]
depending on whether the detectors follow parallel or anti-parallel
accelerated trajectories. Therefore, the dominant frequency scale seen in the
detector's proper time is not fixed, but is effectively shifted by the
acceleration. 

The interplay between acceleration and compactification can be understood in terms of how these two effects modify the balance between the nonlocal correlation harvested from the field and the local noise experienced by the detectors. For small compactification length $L$, the image contributions from the quotient direction are stronger because the field correlations receive contributions from closely spaced topological images.  This enhances the nonlocal correlation term M, giving a larger concurrence at small $L$. However, as the acceleration a increases, the detector response becomes more thermal due to the Unruh effect. The local excitation probabilities increase, and the noise term  $P^E$	
 starts to dominate over the nonlocal correlation. Therefore, in the parallel case, the concurrence is largest for small $L$ and low acceleration, and it decreases as either $L$ or $a$ grows~(\ref{fig:con_acc_L}a). In contrast, in the anti-parallel configuration the nonlocal
Wightman function has a different acceleration dependence, involving
$\cosh[a(\tau+\tau')/2]$, and the relative motion of the detectors can enhance
the nonlocal exchange correlations more efficiently (see~\ref{sec:acceleration}). Consequently,
$|X|$ remains larger over a wider high-acceleration region, giving
rise to the stronger concurrence observed in the anti-parallel case, as seen in figure~\ref{fig:con_acc_L}(b).

\subsection{Superposition: enhanced harvesting region}

The symmetric superposition of the two compactified spacetime sectors exhibits a larger entanglement region than either branch individually. The value of concurrence in superposed geometry can be higher, lower or in between the values of the individual compactified sectors, depending on the value of respective compactification lengths. However, as can be seen from~figure~\ref{fig:concurrence}(b) and (d), the superposition enhances the entanglement region especially at high accelerations amplifying the effect discussed in previous subsection and this effect is seen more prominently in parallel acceleration scenario compared to anti-parallel trajectory as highlighted in figure~\ref{fig:area}.  Whereas each individual compactified branch approaches vanishing concurrence by $a\simeq 5$ in our scans, the superposed configuration retains a measurable and comparatively large concurrence even at the largest accelerations plotted. This slower decay is most prominent at low energy gaps, suggesting that interference can preserve cross-correlations in precisely the regime where local noise would otherwise dominate.
\begin{quote}
Superposed geometry enhances entanglement region in high acceleration regime (specially for parallel acceleration scenario).
\end{quote}
Thus, the superposition of two compactified branches does not necessarily maximize the harvested entanglement relative to each branch individually. Instead, interference redistributes the balance between local noise and nonlocal correlations, thereby enlarging the region of parameter space in which harvesting is possible.  At large acceleration (parallel), the detector's transition probability $P^E$ significantly decreases in the superposed geometry as shown previously (see figure3 in ~\cite{goel2024accelerateddetectorsuperposedspacetime}). However, the cross correlation term $X$ adds up the contributions from each branch thereby leading to higher contribution in concurrence $\mathcal{C}=|X|-P^E$. 

\subsection{Parallel vs.\ anti-parallel acceleration}\label{sec:acceleration}

Previous study~\cite{Liu2022} shows that anti-parallel acceleration systematically yields larger entanglement than parallel acceleration for higher value of the acceleration. We report that a similar behavior is maintained for single compactified and superposed geometry too. For parallel acceleration the highest value of concurrence is 0.4 which occurs when $a\to 0$, whereas for anti-parallel acceleration, the highest concurrence is around 0.6 when acceleration is around 5, as seen in figure~\ref{fig:concurrence}(b) and (d). This also illustrates a clear difference in the behavior of entanglement with parallel and anti-parallel acceleration.  
\begin{quote}
Entanglement increases as acceleration increases for anti-parallel scenario in contrast to parallel acceleration scenario.
\end{quote}

In the anti-parallel configuration the fact that the worldlines separate at late proper times is not the controlling effect for harvesting. The nonlocal term governing entanglement is a double integral weighted by the switching functions, so it is dominated by the portion of the trajectories where the interaction has support (for Gaussian switching, a neighbourhood of the switching centre). In this dominant window, anti-parallel uniform acceleration produces a closest-approach geometry whose separation scale decreases with increasing acceleration.

For the standard anti-parallel Rindler worldlines~\eqref{trajectory} the instantaneous Minkowski separation along the acceleration is
\begin{equation}
\Delta x(\tau)=x_A(\tau)-x_B(\tau)=\frac{2}{a}\cosh(a\tau),
\end{equation}
which is minimized at the centre of the switching window,
\begin{equation}
\Delta x_{\min}=\Delta x(0)=\frac{2}{a},
\end{equation}
(with an additional fixed transverse offset $R$ along $y$ axis, the minimal spatial separation becomes $\sqrt{(2/a)^2+R^2}$). Thus, increasing $a$ makes the detectors closer at closest approach inside the time window that contributes most to the integral, though they diverge rapidly for $|\tau|\gg 0$.

Since the Wightman function entering the nonlocal correlator scales schematically as
\begin{equation}
W(x_1(\tau_1),x_2(\tau_2))\sim \frac{1}{(\Delta t-i\epsilon)^2-|\Delta x|^2},
\end{equation}
the reduction of the invariant interval in the dominant region enhances the magnitude of the cross-correlator and therefore increases the nonlocal term $(X)$ responsible for harvesting. In contrast, for parallel acceleration the two worldlines bend in the same direction and no analogous closest-approach enhancement occurs in the region of dominant phase and switching support; consequently $X$ is not boosted in the same way, while acceleration still increases local contributions. This explains why our numerics can show an increase of concurrence with $a$ in the anti-parallel case but not in the parallel case.

\section{Conclusion and Outlook}\label{sec:conclusion}

In this work, we studied entanglement harvesting by two uniformly accelerated Unruh-Dewitt detectors in quotient Rindler spacetime and in symmetric superpositions of two such compactified geometries. We considered both parallel and anti-parallel acceleration, with the detector separation taken perpendicular to both direction of acceleration and the compactified spatial direction, and analyzed how acceleration, topology, and spacetime superposition jointly affect the extracted entanglement. Our results show that this interplay is nontrivial, but also physically transparent when viewed as a competition between the local excitation noise experienced by each detector and the nonlocal field-mediated exchange term responsible for harvested correlations.

We found that, when the detector separation is perpendicular to the direction of acceleration, the harvested entanglement is generally suppressed relative to more favorable geometric arrangements because the detectors remain more widely separated throughout the interaction. Nevertheless, compactification of the spacetime can partially compensate for this suppression. The image contributions in the compactified Wightman function enhance field correlations and thereby increase the concurrence over sizable regions of parameter space. In particular, compactification extends the range of accelerations over which harvesting remains possible, showing that nontrivial topology can play a constructive role even when acceleration tends to increase local noise (transition probability). In this sense, compactification does not merely modify the quantitative amount of harvested entanglement, but qualitatively improves the robustness of harvesting against acceleration. 

Another main result is that coherent superposition of compactified spacetime sectors provides an additional enhancement. Although the concurrence in the superposed geometry need not exceed that of each individual branch point-wise, the superposition redistributes the relative weights of local and nonlocal contributions through interference, leading to a larger overall harvesting region. This effect is especially pronounced in the high-acceleration regime, where the individual branches approach vanishing entanglement while the superposed configuration can still retain a measurable concurrence. Our analysis therefore suggests that quantum superpositions of spacetime backgrounds can leave operational signatures not necessarily by enhancing entanglement uniformly, but by preserving extractable correlations in regimes where they would otherwise vanish. This suggests that entanglement harvesting may provide an operational probe of quantum features of spacetime geometry. 

A robust feature of our results is the persistent advantage of anti-parallel over parallel acceleration, which has been shown already in previous works~\cite{Liu2022,Salton2015,PhysRevD.107.045010} previously in context of entanglement harvesting with accelerated detectors. We found that the larger entanglement for anti-parallel acceleration compared to the parallel one, continues to hold in both compactified and superposed geometries. Moreover, while parallel acceleration tends to reduce harvesting as the acceleration increases, the anti-parallel configuration can instead support enhanced concurrence in the large acceleration regime. This shows that the favorable role of anti-parallel motion survives as a structurally stable feature of the protocol. In this sense, acceleration, topology, and superposition do not act independently: rather, the geometry of the trajectories determines how strongly compactification and interference can amplify the nonlocal correlations responsible for harvesting. 

Present analysis suggests that entanglement harvesting can function as an operational diagnostic of quantum spacetime structure: rather than probing geometry directly, one probes how spacetime topology and its coherent superposition modify the balance between local detector noise and nonlocal field correlations. This perspective may be useful in developing detector-based signatures of quantum geometry in settings where spacetime is not described by a single classical background, for example, black hole in quantum superposition of different masses. Finally, it would be natural to extend the present framework to finite temperature fields, massive or fermionic fields, non-Gaussian switching profiles, and non-perturbative detector dynamics. These generalizations would help determine whether the robustness of harvesting found here persists in more realistic field-theoretic environments and whether topology or superposition assisted harvesting can be exploited as a resource for relativistic quantum information protocols.

\section*{Acknowledegement}
PS thanks Rita John for her early guidance in initiating the final-year project work, of which this paper is a continuation. The research of LH and AC is supported by ‘The Quantum Information Structure of Spacetime’ Project (QISS) via grant $\#$62312 and ‘The WithOut SpaceTime’ Project (WOST) via grant $\#$63683 from the John Templeton Foundation. 

\onecolumngrid

\begin{appendices}
\section{Evaluation of transition probabilities in~Eq.(\ref{PDLifinal}) and~(\ref{PDL1L2final})}\label{App:A}
To derive the detector's transition probability we here use un-twisted fields $\gamma=1$.  We expand \eqref{PDLi} to derive single detector's transition probability in individual compactified spacetime as
\begin{equation}
\begin{split}
    P_D^{L_i} &=\frac{\eta_0^2}{\mathcal{N}} \int_{-\infty}^{\infty} d\tau \int_{-\infty}^{\infty} d\tau' e^{-(\tau^2 +\tau'^2)/ 2\sigma^2} e^{-i\Omega(\tau - \tau')}\bigg[\sum_{m,n=-\infty}^{\infty} \frac{1}{4\pi i}\text{sgn}(\tau-\tau')\delta\left(\sigma_{((}(x-x')\right)-\frac{1}{4\pi^2\sigma_{((}(x,x')}\bigg],
\end{split}
\end{equation}
where $\sigma_{((}(x,x')=\frac{4}{a^2}\sinh^2\left(\frac{a}{2}(\tau-\tau') \right)-(y-y')^2-(z-z'-(m-n)L_i)^2$ using \eqref{GEO}. For single detector we use $y=y'$, $z=z'$ and change variables as $u=\tau + \tau'$ and $v=\tau -\tau'$. After simplifying the integrals using above change of variables ~\cite{goel2024accelerateddetectorsuperposedspacetime}, this can be separated into three terms such as
\begin{equation}                                            
    P_D^{L_i} = \frac{1}{\mathcal{N}}\sum_{m=n} P_R + \sum_{m\neq n}(A_1+A_2).
\end{equation}
where
\begin{equation}
    P_R =\frac{\eta_0^2\sqrt{\pi}\sigma}{4\pi} \left[ -\Omega -\frac{a^2}{2\pi} \int_{0}^{\infty} dv\,\left[  \frac{e^{-\frac{v^2}{4\sigma^2}} \cos(\Omega v)}{\sinh^2(\frac{a}{2}v)}  - \frac{4}{a^2 v^2}\right] \right].
\end{equation}
is the transition probability of an accelerated detector in Rindler space and the other terms are given by
\begin{equation}
    \begin{split}
        &A_1 = \frac{\eta_0^2 \sigma}{4\mathcal{N} \sqrt{\pi} i}  \int_{-\infty}^{\infty} dv \,  e^{\frac{-v^2}{ 4\sigma^2}}  e^{-i\Omega v}  \text{sgn}(v)\delta\left(\frac{4}{a^2}\sinh^2\left(\frac{av}{2} \right) - ((m-n)L_i)^2\right),\\
        &A_2 = - \frac{\eta_0^2}{4 \mathcal{N} \pi^2} \sigma \sqrt{\pi} \int_{-\infty}^{\infty} dv \,  e^{\frac{-v^2}{ 4\sigma^2}}  e^{-i\Omega v} \frac{1}{\left(\frac{4}{a^2}\sinh^2\left(\frac{a}{2}v \right)-((m-n)L_i)^2\right)}.
    \end{split}
\end{equation}
Using the delta distribution properties
\begin{equation}
\label{Di prop}
    \delta(x^2-a^2)=\frac{1}{2|a|}[\delta(x-a)+\delta(x+a)], \,\,\,\,\,\,\,\text{and}\,\,\,\,\,\,\,\delta(f(x))=\sum_k\frac{\delta(x-\alpha_k)}{\left|f'(\alpha_k)\right|},
\end{equation} 
where $\alpha_k$ are the zero of the function, we can deduce $A_1$ as 
\begin{equation}
    A_1 = \frac{\eta_0^2 \sigma}{4\mathcal{N} \sqrt{\pi} i}  \int_{-\infty}^{\infty} dv \,  e^{\frac{-v^2}{ 4\sigma^2}}  e^{-i\Omega v}  \text{sgn}(v)\left(\frac{\left[\delta\left(v-\frac{2}{a}\sinh^{-1}\left(\frac{a}{2}(m-n)L_i\right)\right)+\delta\left(v+\frac{2}{a}\sinh^{-1}\left(\frac{a}{2}(m-n)L_i\right)\right)\right]}{2|(m-n)L_i|\sqrt{1+\frac{a^2((m-n)L_i)^2}{4}}}\right).
\end{equation}
Assuming $\frac{2}{a}\sinh^{-1}\left(\frac{a}{2}(m-n)L_i\right)=d_i$, we have
\begin{equation}
    A_1 = -\frac{\eta_0^2 \sigma}{4\mathcal{N} \sqrt{\pi} }  \left(\frac{e^{\frac{-d_i^2}{ 4\sigma^2}}  \sin{(\Omega d_i)}}{(m-n)L_i\sqrt{1+\frac{a^2((m-n)L_i)^2}{4}}}\right)\,,
\end{equation}
which yields
\begin{equation}
\begin{split}
    P_D^{L_i} &= \sum_{m=n}\frac{\sqrt{\pi}\sigma}{\mathcal{N}}\frac{\eta_0^2}{4\pi} \left[ -\Omega -\frac{a^2}{2\pi} \int_{0}^{\infty} dv\,\left[  \frac{e^{-\frac{v^2}{4\sigma^2}} \cos(\Omega v)}{\sinh^2(\frac{a}{2}v)}  - \frac{4}{a^2 v^2}\right] \right] -\\ 
    &\sum_{m\ne n}\frac{\eta_0^2 \sigma}{2\mathcal{N} \sqrt{\pi} }\Bigg[\Bigg(\frac{\e^{\frac{-d_i^2}{ 4\sigma^2}}  \sin{(\Omega d_i)}}{(m-n)L_i\sqrt{1+\frac{a^2((m-n)L_i)^2}{4}}}\Bigg) + \frac{1}{2\pi}   \int_{-\infty}^{\infty} dv \, \frac{e^{\frac{-v^2}{ 4\sigma^2}}  e^{-i\Omega v} }{\left(\frac{4}{a^2}\sinh^2\left(\frac{a}{2}v \right)-((m-n)L_i)^2\right)}\Bigg].
\end{split}
\end{equation}
\medskip
Next we evaluate the cross term $P_D^{L_1L_2}$ in \eqref{PDE}. We can expand \eqref{PDL1L2} using \eqref{GEO} to get
\begin{equation}
\begin{split}
    P_D^{L_1L_2} &=\frac{\eta_0^2}{\mathcal{N}}\int_{-\infty}^{\infty} d\tau \int_{-\infty}^{\infty} d\tau' e^{-(\tau^2+\tau'^2)/ 2\sigma^2} e^{-i\Omega(\tau - \tau')}\\
    & \sum_{m,n=-\infty}^{\infty} \frac{1}{4\pi i}\text{sgn}(\tau-\tau')\delta\left(\frac{4}{a^2}\sinh^2\left(\frac{a}{2}(\tau-\tau') \right) -(y-y')^2 -(z-z'+nL_1-mL_2)^2\right)\\
    &-\frac{1}{{4\pi^2}({\frac{4}{a^2}\sinh^2\left(\frac{a}{2}(\tau-\tau') \right)-(y-y')^2-(z-z'+nL_1-mL_2)^2)}}\,.
\end{split}
\end{equation}
Applying the single detector condition $y=y'$, $z=z'$ and considering a variable $K_{L_1L_2}$ such that $K_{L_1L_2}^2=(nL_1-mL_2)^2$ this can be compared with the calculation of $P_D^{Li}$ to get
\begin{equation}
\begin{split}
    P_D^{L_1L_2} &= \sum_{m,n=0}\frac{\sqrt{\pi}\sigma}{\mathcal{N}}\frac{\eta_0^2}{4\pi} \left[ -\Omega -\frac{a^2}{2\pi} \int_{0}^{\infty} dv\,\left[  \frac{e^{-\frac{v^2}{4\sigma^2}} \cos(\Omega v)}{\sinh^2(\frac{a}{2}v)}  - \frac{4}{a^2 v^2}\right] \right] \nonumber \\ 
        &-\sum_{m,n\ne0}\frac{\gamma^{n+m}\eta_0^2 \sigma}{2\mathcal{N} \sqrt{\pi}}\Bigg[ \Bigg(\frac{e^{\frac{-d_K^2}{ 4\sigma^2}}  \sin{(\Omega d_K)}}{K_{L_1L_2}\sqrt{1+\frac{a^2(K_{L_1L_2})^2}{4}}}\Bigg) + \frac{1}{2 \pi}  \int_{-\infty}^{\infty} dv \, \frac{e^{\frac{-v^2}{ 4\sigma^2}}  e^{-i\Omega v}}{\left(\frac{4}{a^2}\sinh^2\left(\frac{a}{2}v \right)-K_{L_1L_2}^2\right)}\Bigg]\,,
\end{split}
\end{equation}
where $d_K=\frac{2}{a}\sinh^{-1}\left(\frac{aK_{L_1L_2}}{2}\right)$, $K_{L_1L_2}^2=(nL_1-mL_2)^2$.

\section{Evaluation of cross correlation term in~Eq.(\ref{Xpara}) and (\ref{xantipara})}\label{App:B}

For \textit{parallel acceleration} we can rewrite \eqref{X} as
\begin{equation}
    X_{((} = - \lambda^2 \Big[ (a^2 + ab) H_1+ (b^2 + ab) H_2 \Big].
    \label{XFor}
\end{equation}
Applying the constant separation between detectors $y-y'=R$, $z=z'$ and the change of variables $u=\tau + \tau'$ and $v=\tau -\tau'$, \eqref{XFor} yields
\begin{equation}
    \begin{split}
        H_1 &= \frac{1}{2} \int_{-\infty}^{\infty} du \int_{0}^{\infty} dv \, \eta_0^2 e^{-(u^2+v^2) / 4\sigma^2}  e^{-i\Omega u} \\
        &\frac{1}{\mathcal{N}}\sum_{m,n=-\infty}^{\infty} \frac{1}{4\pi i}\text{sgn}(v)\delta\left(\frac{4}{a^2}\sinh^2\left(\frac{a}{2}(v) \right) -R^2 -((m-n)L_1)^2\right)-\frac{1}{{4\pi^2}({\frac{4}{a^2}\sinh^2\left(\frac{a}{2}(v) \right)-R^2- ((m-n)L_1)^2)}}.
    \end{split}
\end{equation}
Solving the Gaussian integral in $u$ we have
\begin{equation}
    \begin{split}
        H_1 &= \eta_0^2 \sqrt{\pi}\,\sigma \, e^{-\sigma^{2}\Omega^{2}} \int_{0}^{\infty} dv \,  e^{-v^2 / 4\sigma^2}  \\
        &\frac{1}{\mathcal{N}}\sum_{m,n=-\infty}^{\infty} \frac{1}{4\pi i}\text{sgn}(v)\delta\left(\frac{4}{a^2}\sinh^2\left(\frac{a}{2}(v) \right) -R^2 -((m-n)L_1)^2\right)-\frac{1}{{4\pi^2}({\frac{4}{a^2}\sinh^2\left(\frac{a}{2}(v) \right)-R^2- ((m-n)L_1)^2)}}.
    \end{split}
\end{equation}
We can further divide $H_1$ into two integrals such that 
\begin{equation}
    \begin{split}
        H_1 &= \frac{\eta_0^2 \sqrt{\pi}\,\sigma e^{-\sigma^{2}\Omega^{2}}}{\mathcal{N}} \sum_{m,n=-\infty}^{\infty}(R_1 - R_2).
    \end{split}
\end{equation}
On using \eqref{Di prop} to solve $R_1$ and applying $D_1=(R^2+ (m-n)^2L_1^2)^{\frac{1}{2}}$ we get
\begin{equation}
    R_1 = \frac{\exp\left[{\frac{-\left(\frac{2}{a}\sinh^{-1}\left(\frac{D_1a}{2}\right)\right)^2}{4\sigma^2}}\right]}{8\pi iD_1 \, \, \sqrt{1+\left(\frac{D_1a}{2}\right)^2}},         \,\,\,\,\,\,\,\,\,\,\,\,\,\,\,\,\,\,\,
    R_2 = \frac{1}{{4\pi^2}}\int_{0}^{\infty} dv \, \frac{e^{-v^2 / 4\sigma^2}}{\left(\frac{4}{a^2}\sinh^2\left(\frac{av}{2} \right)-D_1^2\right)}.
\end{equation}
Therefore
\begin{equation}
    \begin{split}
        H_1 &= \frac{\eta_0^2 \sqrt{\pi}\,\sigma e^{-\sigma^{2}\Omega^{2}}}{\mathcal{N}} \sum_{m,n=-\infty}^{\infty}\left(\frac{\exp{\Big[\frac{-\left(\sinh^{-1}\left(\frac{D_1a}{2}\right)\right)^2}{a^2\sigma^2}}\Big]}{8\pi iD_1 \,\, \sqrt{1+\left(\frac{D_1a}{2}\right)^2}}  - \frac{1}{{4\pi^2}}\int_{0}^{\infty} dv \,  \frac{e^{-v^2 / 4\sigma^2}}{\left(\frac{4}{a^2}\sinh^2\left(\frac{av}{2} \right)-D_1^2\right)}\right).
    \end{split}
\end{equation}
Similarly $H_2$ can be simplified to get the final form of $X_{((}$ as
\begin{align}
    X_{((} = &-2 \lambda^2 \eta_0^2 \sqrt{\pi}\,\sigma \, e^{-\sigma^{2}\Omega^{2}} \Bigg[ (a^2 + ab)   \sum_{m,n=-\infty}^{\infty}\Bigg(- \frac{\exp{\Big[\frac{-\left(\sinh^{-1}\left(\frac{D_1a}{2}\right)\right)^2}{a^2\sigma^2}}\Big]}{8\pi iD_1 \,\, \sqrt{1+\left(\frac{D_1a}{2}\right)^2}}  - \frac{1}{{4\pi^2}}\int_{-\infty}^{0} dv \,   \frac{e^{-v^2 / 4\sigma^2}}{\left(\frac{4}{a^2}\sinh^2\left(\frac{av}{2} \right)-D_1^2\right)}\Bigg) \nonumber \\
    &+ (b^2 + ab) \sum_{m,n=-\infty}^{\infty}\Bigg(- \frac{\exp{\Big[\frac{-\left(\sinh^{-1}\left(\frac{D_2a}{2}\right)\right)^2}{a^2\sigma^2}}\Big]}{8\pi iD_2 \,\, \sqrt{1+\left(\frac{D_2a}{2}\right)^2}}  - \frac{1}{{4\pi^2}}\int_{-\infty}^{0} dv \,   \frac{e^{-v^2 / 4\sigma^2}}{\left(\frac{4}{a^2}\sinh^2\left(\frac{av}{2} \right)-D_2^2\right)}\Bigg) \Bigg].
\end{align}
where $D_i=\sqrt{R^2+(m-n)^2L_i^2}$ for $i=1,2$.\\
\medskip

For \textit{anti-parallel acceleration} we follow the calculation for parallel acceleration $X_{)(}$ till \eqref{XFor} and apply the formula~\eqref{GEO2} for anti-parallel Wightman function so that we have $H_1$ as
\begin{equation}
    H_1 = \frac{\eta_0^2}{\mathcal{N}} \int_{-\infty}^{\infty} d\tau \int_{-\infty}^{\tau} d\tau' \sum_{m,n=-\infty}^{\infty} \frac{e^{-\tau^2 / 2\sigma^2}  e^{-\tau'^2 / 2\sigma^2} e^{-i\Omega(\tau + \tau')}}{{4\pi^2}({\frac{4}{a^2}\cosh^2\left(\frac{a}{2}(\tau+\tau') \right)+(y-y')^2+(z-z'-(m-n)L_1)^2)}}.
\end{equation}
Applying the detector separation condition $y-y'=R$, $z=z'$ and using $u=\tau + \tau'$ and $v=\tau -\tau'$, we find
\begin{equation}
    H_1 = \frac{\eta_0^2}{\mathcal{N}} \frac{1}{2}\int_{-\infty}^{\infty} du \int_{0}^{\infty} dv \, 
    \sum_{m,n=-\infty}^{\infty}\frac{\eta_0^2 e^{-(u^2+v^2) / 4\sigma^2}  e^{-i\Omega u} \,\, }{{4\pi^2}({\frac{4}{a^2}\cosh^2\left(\frac{a}{2}(u) \right)+R^2+ ((m-n)L_1)^2)}}.
\end{equation}
Using the Gaussian integral, we can simplify $H_1$ to
\begin{equation}
    H_1 = \frac{\eta_0^2\sigma\sqrt{\pi}}{{8\pi^2}\mathcal{N}} \sum_{m,n=-\infty}^{\infty}   \int_{-\infty}^{\infty} du  \, \frac{e^{-u^2 / 4\sigma^2}  e^{-i\Omega u}}{{\frac{4}{a^2}\cosh^2\left(\frac{a}{2}(u) \right)+R^2+ ((m-n)L_1)^2}},
\end{equation}
and similarly 
\begin{equation}
    H_2 = \frac{\eta_0^2\sigma\sqrt{\pi}}{{8\pi^2}\mathcal{N}} \sum_{m,n=-\infty}^{\infty}   \int_{-\infty}^{\infty} du  \, \frac{e^{-u^2 / 4\sigma^2}  e^{-i\Omega u}}{{\frac{4}{a^2}\cosh^2\left(\frac{a}{2}(u) \right)+R^2+ ((m-n)L_2)^2}}.
\end{equation}
Therefore, we find
\begin{align}
    X_{)(} = -\frac{\lambda^2\eta_0^2\sigma\sqrt{\pi}}{{8\pi^2}\mathcal{N}}   \sum_{m,n=-\infty}^{\infty}\Bigg[ (a^2 + ab) \int_{-\infty}^{\infty} du  \, \frac{e^{-u^2 / 4\sigma^2}  e^{-i\Omega u}}{{\frac{4}{a^2}\cosh^2\left(\frac{a}{2}(u) \right)+D_1^2}} + (b^2 + ab)\int_{-\infty}^{\infty} du  \, \frac{e^{-u^2 / 4\sigma^2}  e^{-i\Omega u}}{{\frac{4}{a^2}\cosh^2\left(\frac{a}{2}(u) \right)+D_2^2}} \Bigg],
\end{align}
where $D_i=\sqrt{R^2+(m-n)^2L_i^2}$ \quad for $i=1,2$.
\end{appendices}
\twocolumngrid
\bibliography{biblio.bib}
\end{document}